\documentstyle[12pt,aps,amssymb,epsfig]{revtex}
\tightenlines
    \renewcommand{\baselinestretch}{1.5}
\begin{document}
\title{The Determination of Microscopic Surface Tension of Liquids with a Curved
       Interphase Boundary by Means of Positron Spectroscopy}
\author{Sergey V. Stepanov, Vsevolod M. Byakov, Olga P. Stepanova  \\
        Institute of Theoretical and Experimental Physics,
        Moscow, 117218, Russia   }
%\date{}
\maketitle
\bigskip
\bigskip
{\bf Abstract:}

\noindent
The method for determination the microscopic surface tension $\sigma(R)$ of
nanobubbles is developed based on the new elaboration of the positronium
bubble model.  In contrast to existing structureless Ps bubble models, our
version contains experimentally known molecular characteristics of liquids.
The relationship, similar to Tolman's equation, between $\sigma(R)$ and the
radius $R$ of the Ps bubble is derived on a microscopic basis. Numerical
values for $\sigma(R)$ are determined for a large number of liquids and
liquified gases. The results are in agreement with the theoretical
expectations and independent evaluations, available in literature.


\bigskip
{\bf The author to whom correspondence should be sent:}

\noindent
Sergey Vsevolodovich Stepanov \\
Institute of Theoretical and Experimental Physics  \\
Bolshaya Cheremushkinskaya 25 \\
Moscow, 117218, Russia \\
\medskip
e-mail: stepanov@vxitep.itep.ru  \\
phone/fax: (095)125-7124; 129-9751

\clearpage

\section{Introduction}

%The problem concerning the influence of curvature of the interphase boundary
%on the surface tension is related to a wide range of physico-chemical
%phenomena.  It is very important in nucleation theory especially for
%estimation of a nucleation rate, production of ultra-dispersion systems, in
%applications capillary problems, wetness, impregnation of membranes etc.

More than one hundred years ago Willard Gibbs in the frameworks of
thermodynamic approach established that if the surface of a liquid is
curved, the surface tension coefficient $\sigma$ gets a function of
curvature $1/r$ of the interphase boundary.  Investigations of Gibbs were
extended by Tolman \cite{Tol49} and others \cite{Hir54,Sch86}.  It was found
that behavior of $\sigma(r)$ for droplets and bubbles is different.  For
droplets $\sigma(r)$ is approximately described by Tolman's equation:

$$
\sigma(r) = \frac{\sigma_\infty}{1+2\Delta/r},
\eqno(1.1)
$$
where $\sigma_\infty$ is the tension for a plane surface and the "Tolman
distance" $\Delta$ is equal to the distance between the equimolecular
dividing surface and the surface of tension.

In case of bubbles $\sigma(r)$ may pass through a maximum \cite{Sch86} and
than goes to zero, but in both cases at small $r$ ~$\sigma(r) \propto r$
\cite{Rus67}.  Special additional assumptions are made concerning
the behavior of the Tolman parameter $\Delta$ as a function of $r$
\cite{Sch86}.

However applicability of the thermodynamic approach to very small systems
such as liquid droplets in the gas phase or bubbles in liquids remains
questionable \cite{Rus67}.  Basic equations hold for systems with large
number of molecules.  Besides it remains unclear how to relate locations of
the dividing surface and the surface of tension to actual positions of
molecules for nanosized voids and droplets. For this we need the
microscopical theory of the surface tension.

The problem is intricate because experimental verification of the
theoretical assumptions made is practically impossible.  Traditional
techniques of surface tension measurements are not sensitive to the
deviations of the microscopic surface tension at small $r$ from the
macroscopic one.  Sometimes the problem is masked by non-equilibrium
properties of the surface formed, which depend on its "age" (so-called
dynamical surface tension).

Solvophobic property of the positronium (Ps) atom, i.e.  its ability to form
a nanobubble in liquid, makes it very attractive probe for investigation of
the possible dependence of $\sigma(r)$ \cite{Bya83,NWJ88}.  Recently we
suggested new approach for determination of the microscopic surface tension
in liquids using positron spectroscopy data \cite{Bya99,Bya00}.  Combining
the data on the ortho-Ps lifetime and width of the "narrow" component of the
ACAR (angular correlation of annihilation radiation) spectrum in the same
liquid it is possible to extract parameters of the Ps bubble. Finally energy
minimization condition allows to determine respective microscopic surface
tension.

In this paper we present detailed description of the Ps bubble model
suggested in \cite{Bya00} and modification of the method of determination of
the microscopic surface tension.

Before starting the explanation of the subject we shell briefly remind
the basis of the positron spectroscopy to introduce terms which will be used
below.

\section{Positron annihilation in matter}

The nonrelativistic approximation in QED gives the following expression for
the spin-averaged probability per second for the annihilation of a
positron-electron pair into two photons, for which the sum of the wave
vectors ${\bf k}={\bf k}_1 + {\bf k}_2$ is within $d^{\sl 3}k$
\cite{Mog95,Akh81}

$$
d\lambda =
    \pi r_0^2 c \cdot \rho_{2\gamma}({\bf k}){d^{\sl 3} k \over(2\pi)^3}.
\eqno(2.1)
$$
Here $r_0=e^2/mc^2$ is the classical electron radius and $c$ is the velocity
of light. The photon-pair momentum density is

$$
\rho_{2\gamma}({\bf k}) =
    \left| \int \phi_{+-}({\bf r},{\bf r}) \cdot
                           e^{-i {\bf k}{\bf r}} d^{\sl 3}r  \right|^2,
\eqno(2.2)
$$
where $\phi_{+-}({\bf r}_+,{\bf r}_-)$ is the wave function of the
annihilating $e^+$-$e^-$ pair.  To obtain total annihilation rate of the
positron we should integrate Eq.(2.1) over the wave vector ${\bf k}$, which
gives

$$
\int \rho_{2\gamma}({\bf k}) {d^{\sl 3}k \over (2\pi)^3} =
     \int \left| \phi_{+-}({\bf r},{\bf r}) \right|^2 d^{\sl 3}r.
\eqno(2.3)
$$

In the case of intrinsic $2\gamma$-annihilation of the Ps atom the wave
function of the annihilating $e^+$-$e^-$ pair may be written as a product

$$
\phi_{+-}({\bf r_+},{\bf r_-}) =
    \psi({\bf r}) \cdot \psi_{\rm Ps}({\bf r_+}-{\bf r_-}),
\eqno(2.4)
$$
where $\psi({\bf r})$ is the function of the center-of-mass coordinate ${\bf
r}=({\bf r_+}+{\bf r_-})/2$ and $\psi_{\rm Ps}(r)$ depends on the
relative coordinate ${\bf r_+}-{\bf r_-}$.  Fourier transform of Eq.(2.4)
and integration over ${\bf k}$ in case when $\psi_{\rm Ps}(r) = e^{-r/2a_B}
/\sqrt{8\pi a_B^3}$ is the ground state of the Ps ($a_B = \hbar /me^2$ is
the Bohr radius) gives the following expressions for the photon-pair momentum
density:

$$
\rho_{2\gamma}({\bf k}) = | \psi_{\rm Ps}(0) |^2
    \left| \int \psi({\bf r})e^{-i {\bf k}{\bf r}} d^{\sl 3}r \right|^2,
\qquad
\int \rho_{2\gamma}({\bf k}) {d^{\sl 3}k \over (2\pi)^3} =
    | \psi_{\rm Ps}(0) |^2 = {1 \over 8\pi a_B^3}.
\eqno(2.5)
$$
From Eq.(2.5) and Eq.(2.1) we obtain the spin-averaged decay rate of the Ps
atom $\lambda_{2\gamma}=r_0^2 c/8a_B^3$. It is equal to one fourth of the
decay rate of the para-positronium $\lambda_{\rm p-Ps} = r_0^2 c/2a_B^3$,
$\lambda^{-1}_{\rm p-Ps} = 123$ ps.\footnote{Account of the higher order
corrections increases the p-Ps lifetime up to 125.2 ps.}

In case of $e^+$ annihilation in matter $\phi_{+-}({\bf r_+},{\bf r_-})$
should be replaced by a total positron and $N$-electron wave function.  In
the frameworks of the independent particle model when all $e^+$-$e^-$
correlations are neglected $\phi_{+-}$ can be approximately written as a
product of the wave functions of the particles involved: $\phi_+({\bf r_+})
\sum_j \phi_-^{(j)}({\bf r}_j)$. Then $\rho_{2\gamma}({\bf k})$ takes the
form:

$$
\rho_{2\gamma}({\bf k}) =
    \sum_j \left| \int \phi_+({\bf r})\phi_-^{(j)}({\bf r})
                           e^{-i {\bf k}{\bf r}} d^{\sl 3} r  \right|^2.
\eqno(2.6)
$$
Here $\phi_+$ and $\phi_-^{(j)}$ are the unperturbed positron and electron
wave functions, respectively, and the sum is taken over all occupied
electron states.  Integration of Eq.(2.6) over ${\bf k}$ gives

$$
\int \rho_{2\gamma}({\bf k}) {d^{\sl 3}k \over (2\pi)^3} =
     \sum_j \int |\phi_+({\bf r})|^2 \cdot
                 |\phi_-^{(j)}({\bf r})|^2 d^{\sl 3}r.
\eqno(2.7)
$$
Usually one assumes that

$$
\sum_j |\phi_-^{(j)}({\bf r})|^2 \approx Z_{\rm eff} n,
    \qquad
\int |\phi_+({\bf r})|^2 d^{\sl 3}r = 1,
\eqno(2.8)
$$
where $n$ is the number density and $Z_{\rm eff}$ is the effective number of
electrons per molecule capable to annihilate with the positron.  Slow
positron can not penetrate deep inside an atom, so core electrons do not
contribute to $Z_{\rm eff}$. So $Z_{\rm eff}$ is close to the number of the
valence electrons.  Substituting Eq.(2.8) into Eq.(2.7) and Eq.(2.1), we
obtain Sommerfeld's result for annihilation rate of "free" positrons in
matter

$$
\lambda_{e^+} = \pi r_0^2 c Z_{\rm eff} n.
\eqno(2.9)
$$

\section{ Ps bubble model}
\subsection{Historical outlook}

In 1957 Ferrel \cite{Fer57} suggested positronium in a liquid creates a
nanocavity (bubble), repelling neighboring molecules outward. It happens
because of a strong exchange repulsion between the electron, composing Ps
atom, and electrons of host molecules. It was the onset of the Ps
bubble model. Its two main aims are the calculation of the lifetime of
the ortho-positronium (o-Ps) and calculation of the shape of the angular
correlation of annihilation radiation (ACAR) spectrum, strictly speaking its
"narrow component", which corresponds to intrinsic $2\gamma$-annihilation of
para-positronium (p-Ps).

Residence of the p-Ps in a bubble practically does not change its lifetime.
It is too short because of prompt intrinsic 2$\gamma$-annihilation.  The
situation is very different for the o-Ps state.  Because of a restriction
imposed by angular momentum conservation 2$\gamma$-decay mode is forbidden
for o-Ps. Thus its most probable decay channel in vacuum is
3$\gamma$-annihilation.  That is why o-Ps lifetime gets approximately 1000
times as large than of p-Ps one and constitutes 142 ns in vacuum.  However
in matter o-Ps may participate in another annihilation process.  It is
so-called pick-off process when e$^+$ annihilates into 2$\gamma$ with an
electron of the opposite spin, belonging to surrounding molecules.  Usually
pick-off annihilation shortens o-Ps lifetime down to several nanoseconds.

To account the pick-off annihilation process Tao modified Ferrel's model
suggesting existence of an electronic layer inside the well, close to its
boundary.  Within the model, which uses an infinite spherically symmetrical
potential well for simulation of the Ps bubble, introduction of such a layer
is a unique way to account for the overlapping of the positron and external
electrons, which is responsible for the pick-off process. Because of
simplicity and physical transparency this model became very popular
\cite{Nak88}.

By the end of the 50's Stewart and Briscoe \cite{Ste59} and Roellig
\cite{Roe67} introduced potential well of a finite height for more adequate
simulation of the trapping potential of the Ps bubble.  Their model came to
present time practically without modifications.  Its basis is the following
\cite{Nak88}.  Ps atom is considered as a point quantum particle, which is
self-trapped in a spherical free-volume cavity.  Action of the surrounding
molecules on the Ps is taken into account via an external potential, which
is simulated by a spherical rectangular potential well with the depth $U$
and radius $R_U$.  A liquid is considered as a structureless continuum.
Almost in all cases the energy of bubble formation is reduced to the surface
energy, which is attributed to the interphase boundary. It is important that
the position of this boundary is also associated with the location of
potential well, i.e. with $R_U$.  So the bubble formation energy is written
as $4\pi R_U^2 \sigma_\infty$.  In calculation of the pick-off annihilation
rate it is assumed that external host electrons do not presented inside the
potential well and o-Ps pick-off annihilation proceeds only due to
overlapping of the o-Ps wave function with the electrons of a medium outside
the boundary of the well.

\subsection{Smooth potentials and concept of average density}

Recently in \cite{Muk97,Muk98,Gan99} the smooth potentials like $U\tanh ^2
(r/R_U)$ and $U(1-e^{-r/R_U})$ were used instead of the sharp finite well
potential.  With this the authors tried to take into account the smooth
variation of the density of medium from the center of the bubble towards
the bulk of the liquid.  Application of these potentials is based on a
possibility to obtain analytical expression for the Ps wave function and the
energy of the Ps ground state.

However we think that this concept of the "average" density profile in the
problem of positronium formation is not well justified. One may admit that
during the bubble formation stage the isolated molecules (so to say a "vapor
phase") may exists inside a "pre-bubble". But we can not take their presence
into account in terms of smooth density distribution (and smooth potential).
Ps motion is much faster then that of molecules and Ps wave function easily
tunes up to current positions of molecules. Ps wave function goes to zero
inside the molecules because of strong exchange and coulombic repulsion.
Such a behavior of the wave function increases Ps kinetic energy (zero-point
energy) and finally promotes pushing all the molecules out of the cavity to
the boundary of the Ps bubble. In contrast to empty bubbles (no Ps inside),
the equilibrium Ps bubble does not contain the "vapor phase".  Presence of
the light quantum particle in the bubble leads to the much more abrupt
density profile on the boundary of the Ps bubble.  Only in this case the
usage of the bell-like Ps wave function (similar to that in Eq.(3.1)) is
meaningful.  We conclude that structure of the interphase region of the Ps
bubble is different from that of usual vapor-liquid boundary.

Obviously smooth Ps wave function is not good approximation outside the
bubble.  Self-consistent results should correspond to a rather small
penetration of the Ps wave function to the bulk of a liquid. As we shell see
below the results obtained on the basis of the present model are
well-matched with this requirement.

Of course the Woods-Saxon potential is suitable for smoothing the sharp edge
of the potential of the rectangular wall \cite{Muk99}.  In the most
realistic case $a \ll R_U$ ~($a$ is the third parameter (beyond $R_U$ and
$U$) entering the Woods-Saxon function) the potential approaches to the
square well shape and the results of the fitting of experimental data using
these two potentials should be similar.

\subsection{General comments about modifications of the Ps bubble model}

Basing on the above comments, we adopt here the potential of the finite
spherically-symmetrical rectangular well for simulation of the Ps bubble, but
introduce additional specifications to make the bubble model more realistic.

As we have seen in the standard model \cite{Nak88,Ste59,Roe67} $R_U$
is overloaded by different physical meanings.  It determines the position of
the potential well, which confines Ps in the bubble. At the same time $R_U$
determines the bubble formation energy, $4\pi R_U^2 \sigma_\infty$. $R_U$
is also related to the pick-off annihilation rate of Ps. It is clear that
description of these effects having such a different physical nature by
means of one parameter is very crude. So we attempt to split
theses effects, introducing additional parameters. Below we reserve for
$R_U$ the meaning of the position of the potential wall only (Fig.1).

How do calculate the energy of the bubble formation? It is an important
question for the Ps bubble model. In section 5 we shell see that the naive
estimation $4\pi R_U^2 \sigma_\infty$ for the surface energy contribution is
not correct. Position of the potential wall responsible for "reflection" of
the Ps into the bubble and position of the surface of tension, related to
intermolecular interaction, are different.  They are placed on different
distances from the center of the bubble. Approximating molecules by spheres
interacting with each other by means of "central" forces, it seems
reasonable to identify the surface of tension with the sphere $S_{R+R_{\rm
WS}}$ passing through the centers of molecules residing on the first
molecular layer of the interphase boundary (Fig.1).  Here $R$ is the radius
of a free-volume and $R_{\rm WS}$ is the Wigner-Seitz radius (${4\over 3}\pi
R_{\rm WS}^3 = 1/n$).\footnote{If molecules are not spherical, but rather
elongated, it could be reasonable to approximate them as a sequence of
spherical fragments. In this case ${4\over 3}\pi R_{\rm WS}^3$ takes sense
of the volume of the fragment.} In typical Ps bubbles the difference
between $R$ and $R+R_{\rm WS}$ is important for estimation of the surface
energy.  This problem will be discussed in Section 4 in more details.

To calculate the o-Ps lifetime we need to know distribution of the
electronic density close to the boundary of the Ps bubble.  Depending on $U$
and $R_U$, $e^+$ may penetrate in some extent through the nearest
molecules (bubble boundary) into the bulk of the liquid. At the
same time electrons belonging to the nearest molecules may reside inside the
potential well.  Thus, electronic density profile and that of the potential,
localizing Ps, should not coincide. Our approach explicitly takes into
to account penetration of outer electrons into the bubble  through the
parameter $\delta$ (Fig. 1).  It leads to additional pick-off annihilation
within the layer of the thickness $\delta$ close to $S_{R_U}$-sphere, but
inside it.\footnote{Of course there are some other reasons, which could lead
to deviation between density and potential profiles. For example,
approaching to the boundary of the bubble one may expect small deepening of
the potential due to polarization interaction as was discussed by Chuang and
Tao \cite{Chu73} in application to silicagel powders. We neglect such
effects here.} It is worth noting that this circumstance allows
to reproduce standard infinite potential well bubble model as a limiting
case $U \to \infty$ of our approach.

\subsection{ortho-Ps lifetime}

By the end of bubble formation (when all the molecules of the "vapor phase"
are pushed out to the bubble boundary) the Ps center-of-mass wave function
takes the form

$$
\psi(r) = \sqrt{\frac{\varkappa}{2\pi(1+\varkappa R_U)}} \cdot
         \left\{ \begin{array}{ll}
{\displaystyle {\sin k_U r \over r}},                         & r \le R_U,\\
\\
{\displaystyle {\sin k_U R_U \over r} e^{-\varkappa(r-R_U)}}, & r \ge R_U, \\
                 \end{array} \right.
\eqno(3.1)
$$
where
$$
k_U^2 = 2m_{\rm Ps} E/\hbar^2
    \qquad
\varkappa^2 = 2m_{\rm Ps}(U-E)/\hbar^2 = 2m_{\rm Ps}U/\hbar^2 -k_U^2.
\eqno(3.2)
$$
Here $E$ is the kinetic energy of Ps and $m_{\rm Ps}=2m_e$ is its mass.
Requirement of smoothness of $\psi(r)$ at $r=R_U$ leads to

$$
\varkappa = - k_U \cot k_U R_U, \qquad \pi/2 \le k_UR_U \le \pi.
\eqno(3.3)
$$
Energy spectrum of the Ps in the well can be obtained form this equation.
When $k_UR_U \to \pi/2$ and $\varkappa R_U \to 0$ the Ps ground state
escapes from the potential well, while the limit $k_UR_U \to \pi$ and
$\varkappa R_U \to \infty$ corresponds to the infinite potential well.  The
relationship

$$
U = \frac{\hbar^2 k_U^2}{2m_{\rm Ps} \sin^2 k_UR_U}
\eqno(3.4)
$$
follows from Eq.(3.2) and Eq.(3.3) and will be used below.

The rate of the pick-off annihilation can be roughly estimated approximating
the factor $\sum_j | \phi_-^{(j)}({\bf r}) |^2$ in Eq.(2.7) as $Z_{\rm eff}
n \cdot \vartheta(r>R)$.  Here $\vartheta$-function equals to unity, if
$r>R$, otherwise it is 0.  Than the r.h.s. of Eq.(2.7) reduces to

$$
\sum_j \int | \phi_+({\bf r})       |^2 \cdot
            | \phi_-^{(j)}({\bf r}) |^2 \cdot d^{\sl 3}r \approx
    Z_{\rm eff} n P_R,
\qquad
P_R = \int_{R}^\infty \left| \psi (r) \right|^2 d^{\sl 3}r.
\eqno(3.5)
$$
Here we approximated $\phi_+({\bf r})$ by the Ps center-of-mass wave
function $\psi(r)$.  $P_R$ is a probability to find Ps (and therefore $e^+$)
outside the free-volume sphere $S_R$. Thus we obtain the relationship for
the pick-off annihilation rate of the o-Ps atom \footnote{Strictly speaking
the annihilation rate of $e^+$-$e^-$-pair having zero spin is 4 times as
large, but the number of host electrons which may form such a zero-spin pair
is 4 times less.  Thus, these effects cancel each other.}:

$$
\lambda_{\rm p-off} = \pi r_0^2 c Z_{\rm eff} n P_R.
$$
In small bubbles $\lambda^{-1}_{\rm p-off}$ practically coincides with the
o-Ps lifetime $\tau_{\rm o-Ps}$, but in large bubbles (for example in liquid
He) we should take into account intrinsic 3$\gamma$ decay of the o-Ps, which
proceeds with the rate $\lambda_{3\gamma}= 1/142$ ns$^{-1}$:

$$
\tau^{-1}_{\rm o-Ps} = \lambda_{\rm p-off} + \lambda_{3\gamma}.
\eqno(3.6)
$$

It is convenient to separate $P_R$ it into two parts:

$$
P_R = P_\delta + P_{R_U},
    \qquad
P_\delta = \int_{R}^{R+\delta} \left| \psi (r) \right|^2 d^{\sl 3}r,
    \qquad
P_{R_U} = \int_{R_U}^\infty \left| \psi (r) \right|^2 d^{\sl 3}r,
    \qquad
R_U = R+\delta.
\eqno(3.7)
$$
Straightforward integrations and account of the normalization condition for
$\psi(r)$ give

$$
P_\delta = \frac{k_U\delta -\sin k_U\delta \cdot \cos(2k_U R_U - k_U\delta)}
                   {k_UR_U-\tan k_UR_U},
    \qquad
P_{R_U} = \frac{\sin^2 k_UR_U}{1-k_UR_U \cot k_U R_U}.
\eqno(3.8)
$$
In the limit of the infinite potential well ($k_U R_U \to \pi$, $\varkappa
R_U \to \infty$, $P_{R_U} \to 0$)  Eq.(3.8) for $P_\delta$ is reduced to the
well-known Tao formula for the o-Ps lifetime \cite{Nak88}

$$
\frac{\tau_{\rm o-Ps}^0}{\tau_{\rm o-Ps}} = P_\delta
    = \frac{\delta}{R_U} - \frac{\sin (2\pi \delta/R_U)}{2\pi},
\eqno(3.9)
$$
where $\tau_{\rm o-Ps}^0$ is usually identified with the positronium
spin-averaged lifetime 0.5 ns. The infinite potential well approach is very
popular because of its simplicity. Knowing experimental value of the o-Ps
lifetime and calculating $R_U$ from the ACAR data (Eq.(3.21)), Eq.(3.9) may
give an information about $\delta$.

In the finite potential well model we suggest to define parameter $\delta$
in the following way.  It was noted by Kobayashi \cite{Hir96}, that if in
Eq.(3.9) the free-volume radius $R=R_U-\delta$ tends to zero and
$\delta= 1.66$ \AA, the energy of the Ps ground state gets equal to the Ps
binding energy in vacuum (6.8 eV).  It indicates that Ps may not exist
without the free volume.  Generalization of this hypothesis for the case of
the finite well is the following:  when $R\to 0$ or $R_U\to\delta$, the Ps
bound state escapes from the potential well. It leads to the following
relationship between $U$ and $\delta$:

$$
U = E\left(k_U R_U = k_U \delta = \frac{\pi}{2}\right) =
    \frac{\pi^2 \hbar^2}{8m_{\rm Ps} \delta^2} =
    \frac{\pi^2}{8} {\rm Ry} \left(\frac{a_B}{\delta}\right)^2,
\eqno(3.10)
$$
where Ry$= \frac{\hbar^2}{2ma_B^2}=13.6$ eV.  Of course in an unperturbed
liquid (without the bubble) $e^+$ and $e^-$ remain bound due to the
long-rage Coulombic interaction, but their binding energy will be small and
the average $e^+$-$e^-$ distance becomes larger than intermolecular
distance. It is the quasi-free positronium in matter. Sometimes it is called
as the swollen Ps.  Electron density of the "own" electron on the positron
in such a state is small in comparison with the other electrons. So,
positron annihilation will look like the free $e^+$ annihilation.

Combination of Eq.(3.10) and Eq.(3.4) gives

$$
k_U \delta = \frac{\pi}{2} \sin k_UR_U.
\eqno(3.11)
$$
This relationship essentially simplifies our approach, because now $P_R$
becomes a function of the parameter $k_UR_U$ only. Thus, knowing o-Ps
lifetime we can directly obtain $k_UR_U$. However, an additional information
is needed to obtain $R_U$ and $k_U$ separately. For this purpose
ACAR-spectroscopy data will be used.

\subsection{Narrow component of ACAR spectra}

The distribution of annihilating photons over $k_z$ is measured by means of
the long-slit angular correlation $e^+$ annihilation apparatus.  The most
reliable information about the Ps state in the bubble is obtained from the
shape of the narrow component of the ACAR spectra, which corresponds to
intrinsic annihilation of the p-Ps:

$$
N(k_z) \propto \int_{-\infty}^{+\infty} dk_x \int_{-\infty}^{+\infty} dk_y
        \rho_{2\gamma}({\bf k}).
\eqno(3.12)
$$
Calculation of the Fourier transform of $\psi(r)$ Eq.(3.1) gives the
photon-pair momentum density:

$$
\rho_{2\gamma}(k) \propto
    \left[ {1 \over k_U^2 - k^2} \left( \sin k_U R_U \cos k R_U -
           {k_U  \over k} \cos k_U R_U \sin k R_U \right) + \right.
$$
$$
\left.+{1 \over k^2 + \varkappa^2} \left( \sin k_U R_U \cos k R_U +
           {\varkappa \over k} \sin k_U R_U \right) \right]^2.
\eqno(3.13)
$$
For the infinite potential well it reduces to

$$
\rho_{2\gamma}(k) \propto
           {k_U^2\over k^2} {\sin^2 k R_U \over (k_U^2 - k^2)^2}.
\eqno(3.14)
$$
It is convenient to carry out an integration over $k_x$- and
$k_y$-components of the photon wave vector using the following
transformation, $k^2 = k_z^2 + k_\perp^2$:

$$
\int_{-\infty}^{+\infty} dk_x \int_{-\infty}^{+\infty} dk_y \dots =
      2\pi \int_0^{\infty} k_\perp dk_\perp  \dots =
       \pi \int_0^{\infty} d(k^2-k_z^2)  \dots =
       \pi \int_{k_z}^{\infty} k dk      \dots .
\eqno(3.15)
$$
Integrating expression for $\rho_{2\gamma}(k)$ in such a manner, we obtain
\cite{Ste90}

$$
N(k_z) \propto \int_{k_z}^\infty k \rho_{2\gamma}({\bf k}) dk
       \propto \left[ {\beta \cos\beta + \varkappa R_U \sin \beta  \over
           (k_U^2 R_U^2 - \beta^2) (\varkappa^2 R_U^2 + \beta^2)} \right]^2,
\eqno(3.16)
$$
which in the limit of the infinite well gives

$$
N(k_z) \propto \int_{k_z R_U}^\infty {d\beta \over \beta}
       {\sin^2 \beta  \over (\pi^2 - \beta^2)^2 }.
\eqno(3.17)
$$

The value of the full width at half maximum, $\Theta_{\rm FWHM}$, of the
narrow component of ACAR spectrum (see Eq.(3.16)) can be obtained from the
following integral equation $2N(k_{\rm FWHM}/2) = N(0)$ or

$$
2\int_{\Theta_*}^\infty \varphi(\beta, k_UR_U) d\beta =
 \int_0         ^\infty \varphi(\beta, k_UR_U) d\beta,
\qquad
\varphi(\beta, x) = \frac{1}{\beta}
    \left[ \frac{\beta \cos\beta -x \sin \beta \cot x}
           {(x^2 - \beta^2) (\beta^2 +x^2 \cot^2 x)} \right]^2,
\eqno(3.18)
$$
where
$$
\Theta_* = \frac{k_z^{\rm FWHM}R_U}{2} =
           \frac{m_ec}{\hbar} R_U \frac{\Theta_{\rm FWHM}}{2}.
\eqno(3.19)
$$
Here $m_e c$ is a momentum of one of the annihilating $\gamma$-quanta.

It is important that knowing $k_UR_U$ (from the o-Ps lifetime data) and
$\Theta_{\rm FWHM}$ from ACAR measurements we may obtain $R_U$ and all other
parameters of the Ps trap ($U$ and $\delta$). If $U \to \infty$, in
Eq.(3.18) we should set $x \to \pi$, which leads to

$$
2\int_{\Theta_*}^\infty \frac{d\beta}{\beta}
    \left( \frac{\sin \beta}{\pi^2 - \beta^2} \right)^2 =
 \int_0         ^\infty \frac{d\beta}{\beta}
    \left( \frac{\sin \beta}{\pi^2 - \beta^2} \right)^2
\eqno(3.20)
$$
with the same meaning of $\Theta_*$. Numerical solution of Eq.(3.20) gives
$\Theta_* = 2.1480$. Substituting this value to Eq.(3.19) we obtain simple
relation between $\Theta_{\rm FWHM}$ and radius of the bubble \cite{Nak88}:

$$
R_\infty \hbox{ [\AA]} = \frac{16.65}{\Theta_{\rm FWHM}\hbox{ [mrad]}}.
\eqno(3.21)
$$

Usually for extraction of the narrow component from the total ACAR spectrum
it is decomposed into a set of gaussians. To visualize uncertainty which
comes from neglecting deviation in shape between Eq.(3.16) and the gaussian
with the same $\Theta_{\rm FWHM}$, in Fig.2 we plotted several spectra for
different Ps traps.  For rather "deep" well ($k_UR_U \ge 2.5$) the
difference is not large, but for "shallow" traps ($k_UR_U \le 2.5$)
this deviation should be taken in to account in decomposition of the ACAR
spectrum.

\section{ Elementary model of a cavity formation }

Let   molecules   in  the  liquid  interact,  for  example,  through  to  the
Lennard-Jones potential:

$$
\varphi_{LJ}\left(\frac{\bar r}{r}\right) = \epsilon
    \left[ \left( \frac{\bar r}{r} \right)^{12} -
          2\left( \frac{\bar r}{r} \right)^6      \right].
\eqno(4.1)
$$
When $r$ is equal to average molecular separation $\bar r$ the pair-wise
molecular potential energy reaches its minimum:  $\varphi_{LJ}(r={\bar r}) =
- \epsilon$.  In equilibrium $\bar r$ is related with the number density $n$
as $\bar r \approx n^{-1/3}$. We shall consider molecules as semi-hard
spheres of radius $\bar r/2$ with only nearest neighbor interactions
($\varphi_{LJ}$ approaches zero very rapidly with increasing $r$).

Since the energy required to form a surface arises from the decrease in
number of bonds of molecules at the boundary, the accompanying decrement in
coordination number, $\nu$, needs to be estimated. The bulk liquid we
imagine as a closely packed structure.  Everywhere in it one may find four
nearest molecules, which form a regular tetrahedron of edge length $\bar r$,
whose centers lie on a sphere $S_{r_1}$ of radius $r_1 = \bar r \sqrt{3/8}$
(Fig.3). Therefore, the area per molecule on $S_{r_1}$ is $\frac{1}{4} \cdot
4\pi r_1^2 =\pi r_1^2$ and the coordination number will be

$$
\nu_0 = {4\pi \bar r^2 \over S_1}  =
    4\left({\bar r\over r_1}\right)^2 = {32 \over 3}.
\eqno(4.2)
$$
In what follows we neglect the difference between $r_1$ and the Wigner-Seitz
radius because it may be seen that $r_1 = \left({4\pi\over 3}\right)^{1/3}
\cdot \sqrt{{3\over 8}} R_{\rm WS} = 0.986 R_{\rm WS}$.  Hence for all
practical purposes we may take that the volume of the sphere $S_{r_1}$ of
the closest neighbors to have the same volume as the average volume per
molecule in the liquid or, in other words, $r_1 \approx R_{\rm WS}$.

Next permit the positronium to create a spherical free volume $4\pi R^3/3$
(or bubble) of radius $R$ inside the tetrahedron and thereby pushing the
molecules outward. The other molecules also rearrange themselves to settle
on the first molecular layer (FML) of radius $r_1+R$, Fig.4.
The number $N_{\rm FML}$ of molecules lying on this layer can be
estimated from its area, $4\pi(r_1+R)^2$, dividing by the area per
molecule, $\pi r_1^2$, that is:

$$
N_{\rm FML} = {S_{r_1+R} \over S_1} = 4\left({r_1+R\over r_1}\right)^2.
\eqno(4.3)
$$
Because of formation of the bubble with the free volume $4\pi R^3/3$ the
coordination number which was $\nu_0$ suffers a decrement in proportion to
the free area per molecule residing on the FML, $S_R/N_{\rm FML}$, divided
by the area per molecule in the bulk, $4\pi (\bar r/2)^2$, namely

$$
\Delta \nu = \frac{S_R/N_{\rm FML}}{\pi \bar r^2 } \cdot \nu_0 =
                 4\left({R\over r_1+R}\right)^2,
\qquad     S_R = 4\pi R^2.
\eqno(4.4)
$$
The fraction $\frac{S_R/N_S}{4\pi (\bar r/2)^2 }$ represents a factor
decreasing the coordination number.  Accordingly, the surface energy should
be proportional to the number of broken bonds viz.

$$
E_\sigma \sim N_{\rm FML} \Delta \nu \sim R^2/r_1^2.
\eqno(4.5)
$$
Also in view of the fact that we are dealing with the central forces and
these act at the center of molecules we expect the surface of tension is
located at $R_\sigma=R+r_1 \approx R+R_{\rm WS}$. Thus one would expect the
surface energy to be proportional to the area of the surface of tension,
$S_{R_\sigma} = 4\pi R^2_\sigma$, with the curvature dependent coefficient
of proportionality $\sigma(R_\sigma)$:

$$
E_\sigma \sim R^2_\sigma \sigma(R_\sigma).
\eqno(4.6)
$$
In the limit $R\to \infty$ the surface tension energy $E_\sigma$ should
reproduce standard relationship for the plane surface viz. $E_\sigma = 4\pi
R^2 \sigma_\infty$.  So we reconstruct coefficient of proportionality in
Eq.(4.6). Finally, comparing Eq.(4.5) and Eq.(4.6) we obtain

$$
E_\sigma = 4\pi R_\sigma^2 \sigma(R_\sigma),
\qquad
\frac{\sigma(R_\sigma)}{\sigma_\infty} = \left( {R\over R_\sigma} \right)^2
\approx \frac{1}{(1+ R_{\rm WS}/ R)^2},
\qquad
R_\sigma = R+r_1 \approx R+R_{\rm WS}.
\eqno(4.7)
$$

Furthermore, regarding subsequent minimization of the total energy (balance
condition), we prefer to rewrite the surface energy of the bubble in an
integral representation for convenience and thus introduce the surface
tension function $\tilde\sigma (r)$ through:

$$
E_\sigma = \int {2 \tilde\sigma (r) \over r} d^{\sl 3}r,
$$
which also happens to be the Laplace form. Of course if $\tilde\sigma$
is put equal to $\sigma_\infty$ and integration is performed over
the volume $4\pi R^3/3$, one obtains the result $E_\sigma = 4\pi R^2
\sigma_\infty$. However, here  we must proceed with the space integration
over the range from $R_{\rm WS}$ to $R_\sigma$ and obtain $\tilde\sigma (r)$
by solving the integral equation:

$$
\int_{R_{\rm WS}}^{R_\sigma}
            {2 \tilde\sigma (r) \over r} d^{\sl 3}r =
4\pi R^2 \sigma_\infty.
\eqno(4.8)
$$
Eq.(4.8) being differentiated with respect to $R$, gives us a Tolman-like
expression:

$$
\tilde \sigma(R_\sigma) = \sigma_\infty \frac{R}{R_\sigma}=
                      {\sigma_\infty \over 1+R_{\rm WS}/R}.
\eqno(4.9)
$$
This relationship sheds light on a physical meaning of the Tolman length
$\Delta$ in Eq.(1.1) through the relation $2\Delta \approx R_{\rm WS}$. Thus
from the molecular point of view $\Delta$ accounts for the molecules on the
curved first molecular layer as having more neighbors (less broken bonds)
that the molecules on the plane interphase surface.  It is worth noting that
the surface tension coefficient of the curved boundary depends not only on
the position of the surface of tension, but also on the type of
representation of the surface energy.

\section{ Energy minimization }

Energy minimization condition can be naturally inscribed into our
consideration in the following manner. As we have demonstrated above all
parameters of the Ps trap can be obtained from the o-Ps lifetime and the
width of the narrow component of the ACAR spectrum.  However $\tilde
\sigma$, entering Eq.(4.8), may be considered as an unknown function of $R$,
neglecting the theoretical prediction, Eq.(4.9). Thus we suggest to extract
so-to-say "experimental" dependence of $\tilde \sigma (R)$ from the
principal of the minimum of the total energy $E_{tot}$ of the Ps
bubble\footnote{The usage of the integral representation for $E_\sigma$
allows to avoid an appearance of the derivative of $\tilde\sigma$ over $R$
minimizing $E_{tot}$.}:

$$
E_{tot} = E_\sigma + \frac{4\pi}{3}R^3 p_0 + E =
   \int_{R_{\rm WS}}^{R_\sigma} \frac{2\tilde \sigma(r)}{r} d^{\sl 3}r +
   \frac{4\pi}{3}R^3 p_0 +
   \frac{\hbar^2 k_U^2}{2m_{\rm Ps}}.
\eqno(5.1)
$$
Here we added the term $\frac{4\pi}{3} R^3 p_0$, which is the work against
external pressure $p_0$. This term is important for rather large Ps bubbles
in liquified gases.

To proceed with the minimization of $E_{tot}$ let us first figure out a
useful relationship for $dE/dR$:

$$
\frac{dE}{dR} = \frac{2k_U E}{\tan k_UR_U - k_UR_U},
\eqno(5.2)
$$
which can be obtained from Eq.(2.2) and Eq.(2.3) by differentiation over $R$
and further exclusion of $\varkappa$. Then the balance condition
$dE_{tot}/dR = 0$ may be written as

$$
\frac{d}{dR}
    \left[ \int_{R_{\rm WS}}^{R_\sigma} \frac{2\tilde \sigma(r)}{r} d^{\sl 3}r +
           \frac{4\pi}{3}R^3p_0 + E \right] = 0.
\eqno(5.3)
$$
This equation gives

$$
\frac{\tilde \sigma(R_\sigma)}{\sigma_\infty} =
    \frac{(R_{eq}^\infty)^4}{R_\sigma R_U^3} \cdot
        \frac{k_U^3 R_U^3}{\pi^2(k_UR_U-\tan k_UR_U)} -
    \frac{p_0 R^2}{2\sigma_\infty R_\sigma},
\qquad
    R_{eq}^\infty =
        \left(\frac{\pi a_B^2 {\rm Ry}}{8\sigma_\infty} \right)^{1/4}.
\eqno(5.4)
$$
For the infinite potential well Eq.(5.4) simplifies to:

$$
\frac{\tilde \sigma(R_\sigma)}{\sigma_\infty} =
    \frac{(R_{eq}^\infty)^4}{R_\sigma R_U^3} -
    \frac{p_0 R^2}{2\sigma_\infty R_\sigma}.
\qquad
\eqno(5.5)
$$

\section{Results and discussion}

\subsection{The bubble is a deep potential well for Ps}

Using Eqs.(5.4-5.5) and knowing $\tau_{\rm o-Ps}$ and $\Theta_{\rm FWHM}$
from the $e^+$-$e^-$ annihilation experiments it is possible to obtain
microscopic values of $\tilde\sigma$ for the interphase boundary having
curvature radius about several angstroms.\footnote{Without the last term
with $p_0$ Eq.(5.5) was obtained in \cite{Bya99}.} Deriving these
relationships we did not use any particular expression for  the surface
tension coefficient vs curvature radius of the boundary.  We assumed only
that cavity formation energy is written in the Laplace form, Eq.(4.8).

For all investigated molecular liquids ratios $\tilde \sigma /\sigma_\infty$
are less than unity. Variations between the values corresponding to the
finite well, Eq.(5.4), and the infinite well, Eq.(5.5), is small (Tables 1,
2; Fig.5).  The reason is that the obtained depth $U$ of the potential well
is rather large $U\gg E$. In Section 3 we mentioned that this inequality
ensures self-consistent usage of the bell-like Ps wave function in the
frameworks of the present theory.

In \cite{Muk99} in some liquids (glycerin, ethylene glycol, methanol-water
mixtures) was not fulfilled. It implies large penetration of the Ps to the
bulk of a liquid, which means that the results might not be
reliable.\footnote{We think that in \cite{Muk97,Muk98,Gan99,Muk99}
expression for the pick-off annihilation rate the square of the Ps
psi-function $|\psi(r)|^2$ has to be multiplied on the respective value of
the electronic density and than this product should be integrated over space
variables.}

\subsection{Profiles of the potential well and that of electronic density
are different}

Previous formulations of the Ps bubble model, utilized finite well
potential, were not able to reproduce in a limiting case $U \to \infty$ the
Tao formula (3.9) for $\tau_{\rm o-Ps}$, obtained within the infinite
potential well model.\footnote{In the frameworks of the conventional finite
well model in the limit $U\to \infty$ the Ps wave function is confined
within the bubble only and does not overlap with outer electrons.  Therefore
the pick-off annihilation rate equals to zero.} We avoid this drawback in
the present formulation through the parameter $\delta$,
which accounts some penetration of the outer electrons within the well.
Thus $\delta$ discriminates profiles of the potential and that of electronic
density.

\subsection{Separation of the position of the potential wall and the surface
            of tension}

Another new and important element of the present formulation is the
separation of the position of the potential well ($R_U$), which reflects Ps
into the bubble, and that of surface of tension ($R+R_{\rm WS}$), related to
interaction\footnote{rupture of intermolecular bonds.} between the
molecules residing on the first molecular layer.

\subsection{Correlation between the Tolman length and the Wigner-Seitz
            radius}

Comparison between "experimental" values of $\tilde \sigma /\sigma_\infty$
and respective theoretical prediction, Eq.(5.9), for two values of the
parameter $2\Delta$  ($R_{\rm WS}$ and $3 R_{\rm WS}$) are shown in
Fig.5.   One may conclude that there is reasonable agreement between the
theory and experimental data in spite of many simplifications done, which
could be inadequate especially in polar liquids or in liquids with
large non-spherical molecules.

Assuming validity of the Tolman equation and knowing $\tilde \sigma
/\sigma_\infty$ values, one may calculate the ratios $2\Delta/R_{\rm WS}$.
They are within the interval from 1 to 3, which is in a reasonable agreement
with Eq.(4.9).\footnote{Experimental uncertainty of $2\Delta/R_{\rm WS}$ is
about 100\%.} Our values of $2\Delta$ for water are 5.4 and 3.3 \AA\ (see
Tables 1 and 2), which correlates well with available literature
data\footnote{Data for droplets.}:  6.0 \cite{Ask98}, 1.8 \cite{Akh72} and
2.0 \cite{Tol49} \AA.  The same takes place in liquid argon:  we obtained
$2\Delta = 4.5$-3.8 \AA, while in \cite{Hir54}~ $2\Delta =7.3$ \AA.

In spite of the large uncertainty of the $\Delta/R_{\rm WS}$ values it is
worse noting their correlation within the classes of different chemical
compounds. In isooctane, neopentane and tetramethylsilane, i.e. in liquids
with round molecules values of $2\Delta/R_{\rm WS}$ are close to
unity\footnote{Small value of $2\Delta/R_{\rm WS}$ in 1,4-dioxane, which
does not belong to this class of compounds, probably related to the special
alignment of dioxane molecules on the surface of the bubble.} in agreement
with Eq.(4.9).  In liquid hydrocarbons made up from normal, cyclic and
aromatic molecules, in higher alcohols, diethylether, acetone the ratio
$2\Delta/R_{\rm WS}$ increases up to 1.8 in average. Probably it is related
to the orientation of molecules when their maximal linear dimension
primarily directed perpendicularly to the surface of the Ps bubble.  Higher
values of $2\Delta/R_{\rm WS}$ (up to 2.7) occur in low alcohols, water and
acetonitrile.  These liquids consist of small polar molecules strongly
interacting with each other. It is interesting that $2\Delta/R_{\rm WS}$ in
CS$_2$ is also high in spite of CS$_2$ molecule has no dipole moment.
However, as follows from radio-spectroscopy studies \cite{Sha72}, complicate
polar molecular associates are presented in liquid CS$_2$. As a result, an
effective dipole moment per molecule turns out to be approximately 0.1 D and
intermolecular binding energy between CS$_2$ gets about 0.05 eV.  Thus one
may expect that there is a correlation between $2\Delta/R_{\rm WS}$ and
efficiency of intermolecular interaction. The latter can be characterized,
for example, by critical pressure $p_{cr} = \frac{a}{27b^2}$, where $a$ and
$b$ are known parameters of the Van-der-Waals equation (Fig.6). This
correlation is clearly seen in a homological series of alcohols. It is
difficult to expect better correlation between the data obtained from rather
schematic Ps bubble model and parameters of the Van-der-Waals equation,
applied to the liquid phase.

\subsection{Ps is an electrically neutral probe of the interphase boundary}

A perturbation of the surface caused by the presence of the Ps atom in the
bubble on does not extend deep inside the liquid because of electrical
neutrality of the positronium. One may expect that the structure of the
interphase boundary in the Ps bubble will be more close to the free surface
than in the case of the bubbles formed by excess electrons. It is also worse
mentioning that the surface of the Ps bubble is rather "fresh".  Its age is
no more than some nanoseconds.  Contrary, surfaces studied by means of
conventional methods have the ages many orders higher.  So in liquids with
rather long relaxation times properties of the boundary of the Ps bubble and
that of the equilibrium surface may be different.

\section{Conclusion}

Major part of this work is the development of the Ps bubble model, which
forms the basis of the method for determination if the local surface
tension. The modifications done are the following.

Firstly, it is taken into account, that position $R_U$ of the potential well
does not coincide with the position of the surface of tension.  Secondly, we
admit a possibility of the electrons of the nearest molecules to penetrate
inside the Ps bubble.  Just this feature allows us a to reproduce the Tao
formula as a limiting case $U \to \infty$ of our finite potential well
model. It is important that we did not introduce undefined parameters to the
model. It is due to the additional constrain, Eq.(3.10), which has the
following physical meaning:  in molecular liquids "preexisting" free volume
can not localize Ps atom; in an "unperturbed" liquid Ps exists in the
quasi-free (swollen) state which manifests experimentally like free $e^+$
annihilation.

Elementary "geometric" consideration of the cavity formation, done on a
molecular level, made possible to reproduce the Tolman equation and clear up
the physical sense of the curvature dependence of surface tension. We have
found that Tolman's length $\Delta$ takes into account an increase of the
number of the nearest neighbors of a molecule, residing on the first
molecular layer of the curved boundary of the bubble in comparison with
coordination number of a molecule on flat interphase boundary. It is shown
that particular value of the surface tension coefficient depends on the
position of the surface of tension, and on the choice of explicit expression
for the energy of the bubble formation.

It is shown that the data on o-Ps lifetimes and widths of narrow
component of ACAR spectra allow to obtain "experimental" values of the
surface tension coefficient without any hypotheses about its concrete
functional dependence vs $R$. The results are in a satisfactory agreement
with the Tolman relationship Eq.(4.9) and other independent evaluations of
the surface tension coefficients.  We have found a correlation between
the ratio $2\Delta/R_{\rm WS}$ and critical pressure.  It indicates
on its usefulness in consideration of intermolecular interactions and
structure of surface layers.

\acknowledgments
We thank the Russian Foundation of Basic Research for Grant 98-03-32058a
in support of this work.

%%%%%%%%%%%%%%%%%%%%%%%%%%%%%%%%%%%%%%%%%%%%%%%%%%%%%%%%%%%%%%%%%%%%%%
\clearpage
\centerline{{\Large Figure Captions}}
\bigskip
\bigskip
Figure 1.

\medskip
\noindent
The Ps bubble. The center-of-mass wave function $\psi(r)$ of Ps confined
by a spherical potential well of the depth $U$ and radius $R_U$, Eq.(3.1).
$4\pi R^3/3$ is the free volume. $R+R_{\rm WS}$ is the distance from the
center of the bubble to the centers of molecules residing on the first
molecular layer. The parameter $\delta$ characterizes penetration of host
electrons inside the potential well of the bubble.

\bigskip
Figure 2.

\medskip
\noindent
Different normalized (unit area below lines) narrow components of ACAR
spectra, having the same $\Theta_{\rm FWHM}$.  Circles represent the
gaussian line $\frac{1}{\sqrt{2\pi M_2}} \exp (-\frac{\Theta}{2M_2})$,
$\Theta = \hbar k_z / m_e c$ with the second moment $M_2=\Theta_{\rm
FWHM}^2/(8\ln 2)$.  Solid line represents the narrow component for in the
case of an infinite potential well, Eq.(3.17).  Dashed lines are plotted
according to Eq.(3.16) for different values of $k_U R_U$.

\bigskip
Figure 3.

\medskip
\noindent
Tetrahedron formed by joining the centers of four nearest neighbors which in
turn lie on a sphere of radius $r_1$ (the fourth molecule, nearest to the
reader, is not shown).  Molecules are simulated as spheres of radius $\bar
r/2$ ($\bar r$ being the average intermolecular distance).

\bigskip
Figure 4.

\medskip
\noindent
Creation of the free volume spherical void in a liquid.

\bigskip
Figure 5.

\medskip
\noindent
Dependence of the relative surface tension vs $R/R_{\rm WS}$. Values of
$\tilde \sigma /\sigma_\infty$ for different liquids are represented with a
help of the respective numbers listed in Table 2. Upper line represents
Tolman's equation (4.9) and the curve below is the same relationship but
with the factor of three in the Tolman parameter viz.  $\tilde
\sigma/\sigma_\infty = (1+3R_{\rm WS}/R)^{-1}$.

\bigskip
Figure 6.

\medskip
\noindent
Correlation between $2\Delta/R_{\rm WS}$ and critical pressure $p_{cr}$ in
different liquids at room temperature (Table 2). Linear proportionality
between $\Delta/R_{\rm WS}$ and $p_{cr}$ in homological series of alcohols
is clearly seen (corresponding numbers are encircled).

%%%%%%%%%%%%%%%%%%%%%%%%%%%%%%%%%%%%%%%%%%%%%%%%%%%%%%%%%%%%%%%%%%%%%%
\clearpage
\begin{figure}[t]
\epsfig{file=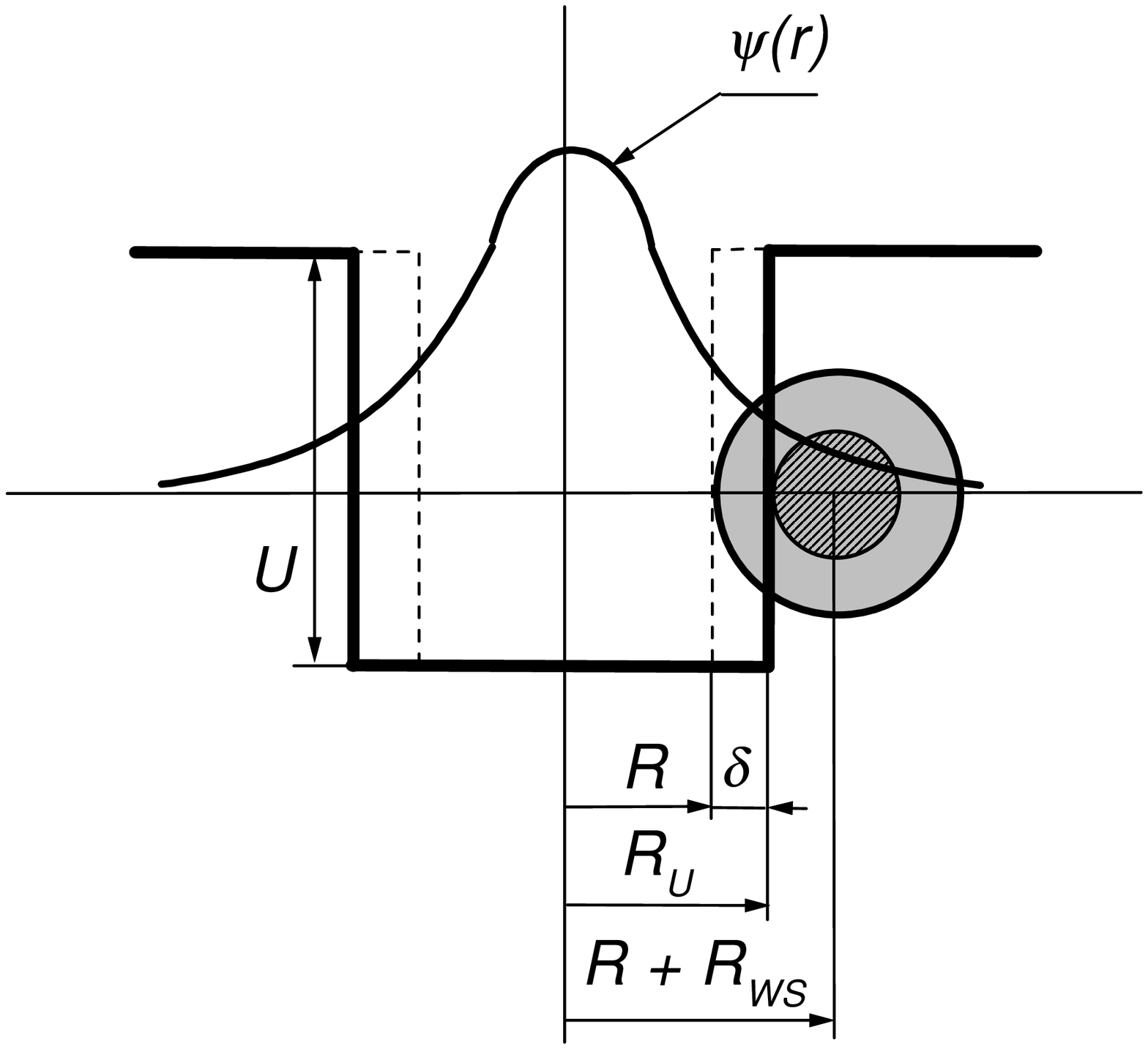, width=15cm}
\vspace{1cm}
\caption{}
\end{figure}

%%%%%%%%%%%%%%%%%%%%%%%%%%%%%%%%%%%%%%%%%%%%%%%%%%%%%%%%%%%%%%%%%%%%%%
\begin{figure}[t]
\epsfig{file=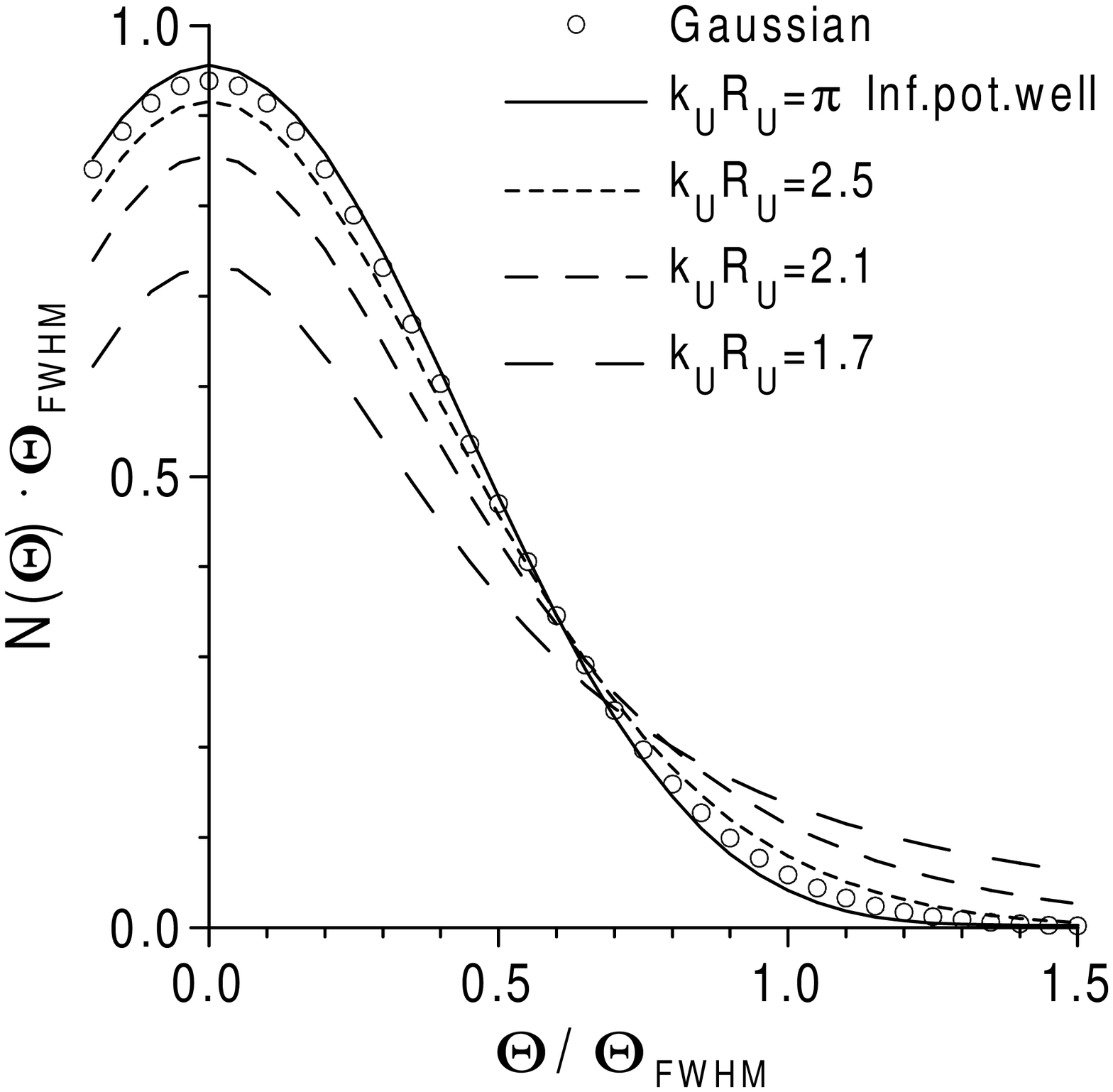, width=17cm}
\vspace{3cm}
\caption{}
\end{figure}

%%%%%%%%%%%%%%%%%%%%%%%%%%%%%%%%%%%%%%%%%%%%%%%%%%%%%%%%%%%%%%%%%%%%%%
\begin{figure}[t]
\epsfig{file=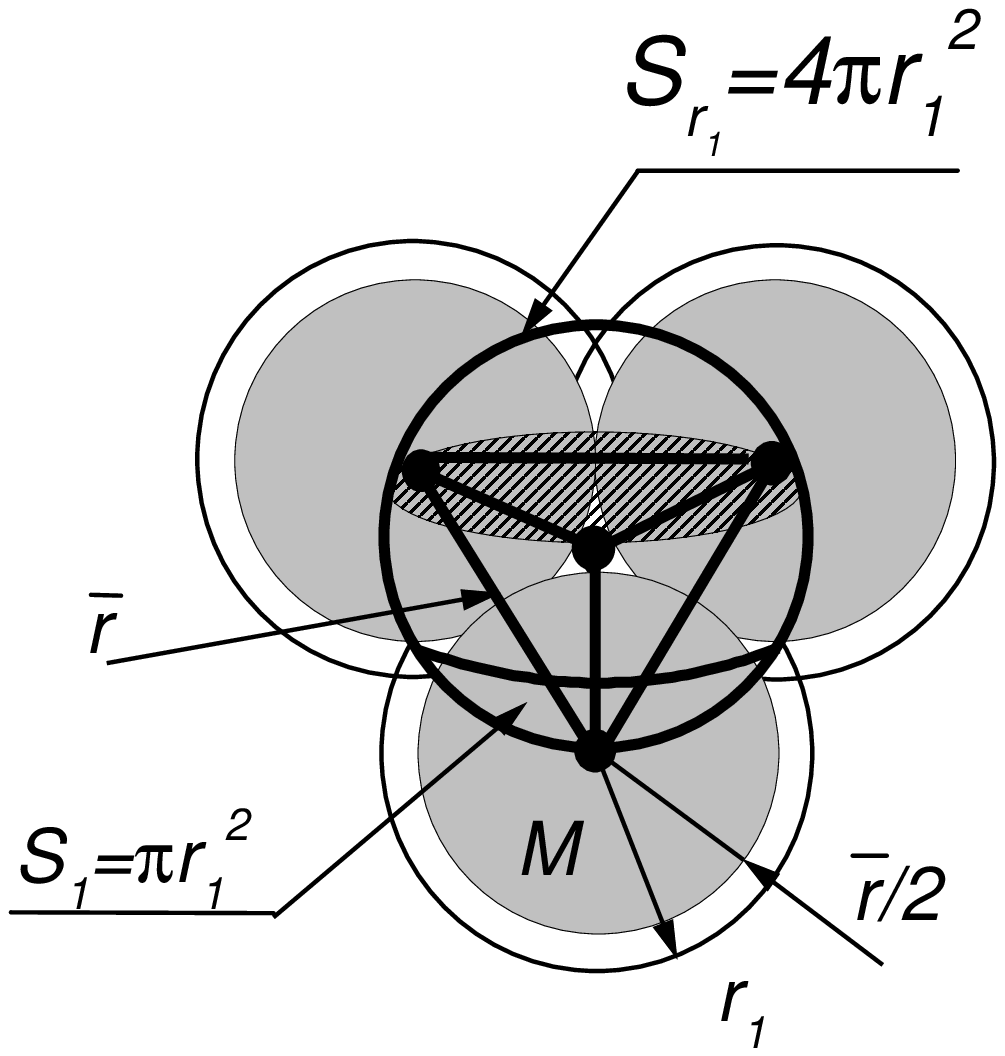, width=17cm}
\vspace{3cm}
\caption{}
\end{figure}

%%%%%%%%%%%%%%%%%%%%%%%%%%%%%%%%%%%%%%%%%%%%%%%%%%%%%%%%%%%%%%%%%%%%%%%
\begin{figure}[t]
\epsfig{file=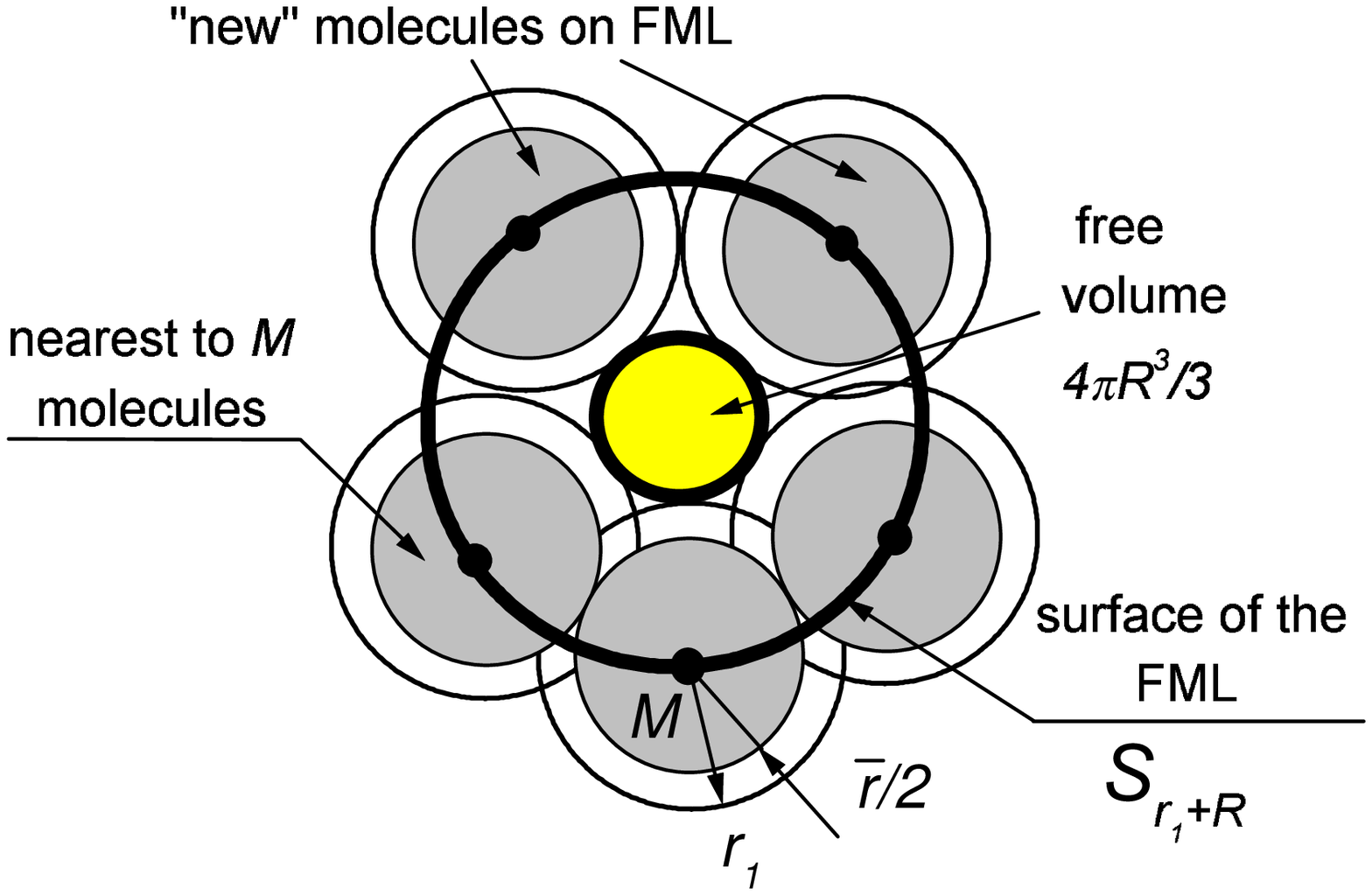, width =17cm}
\vspace{3cm}
\caption{}
\end{figure}

%%%%%%%%%%%%%%%%%%%%%%%%%%%%%%%%%%%%%%%%%%%%%%%%%%%%%%%%%%%%%%%%%%%%%%
\clearpage
\begin{figure}[t]
\epsfig{file=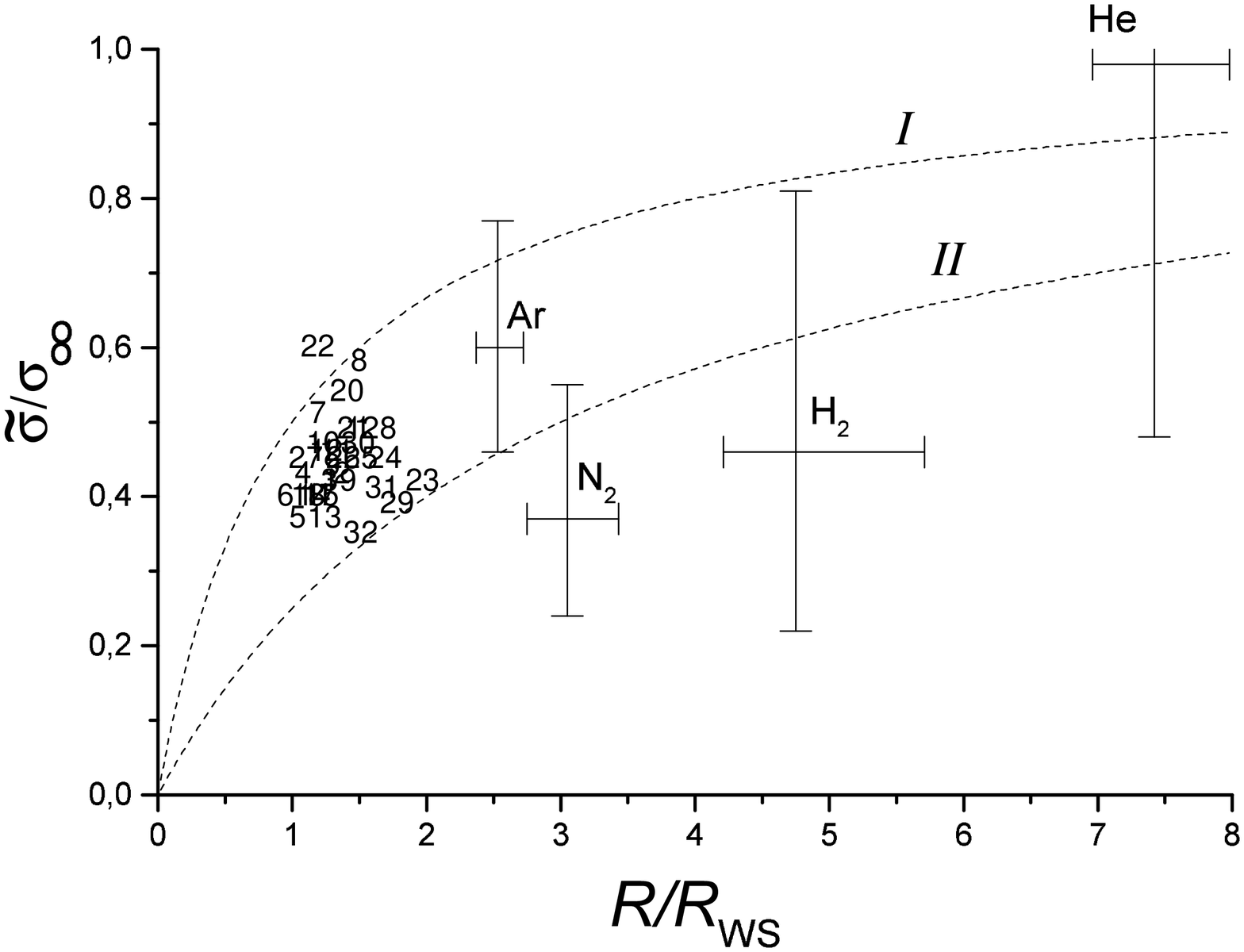, width =17cm}
\vspace{3cm}
\caption{}
\end{figure}

%%%%%%%%%%%%%%%%%%%%%%%%%%%%%%%%%%%%%%%%%%%%%%%%%%%%%%%%%%%%%%%%%%%%%%%
\clearpage
\begin{figure}[t]
\epsfig{file=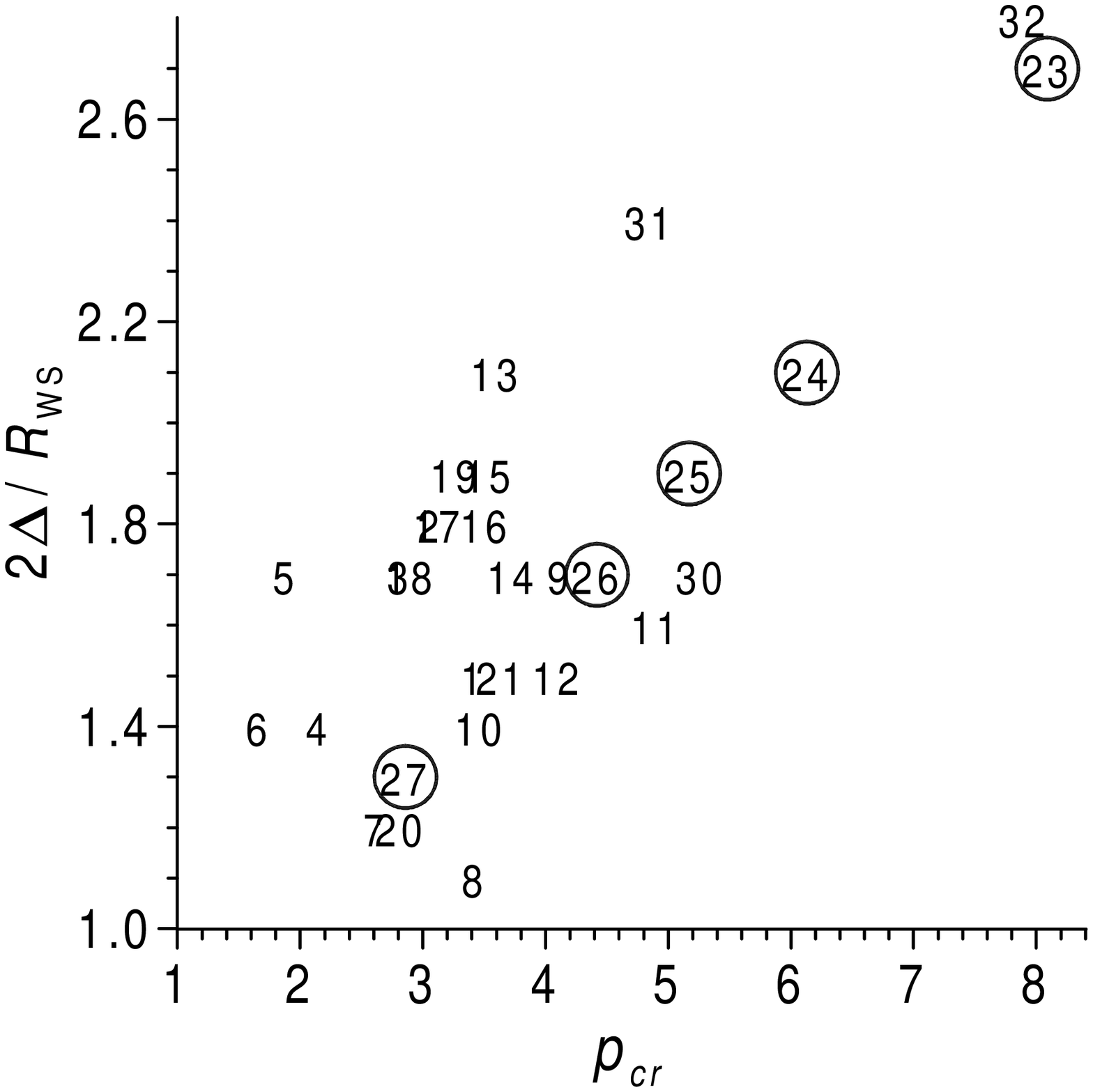, width =17cm}
\vspace{3cm}
\caption{}
\end{figure}

%%%%%%%%%%%%%%%%%%%%%%%%%%%%%%%%%%%%%%%%%%%%%%%%%%%%%%%%%%%%%%%%%%%%%%%
\clearpage
\begin{table*}[t]
\caption{Some parameters of liquids and Ps bubbles, obtained
         within the infinite potential well model.}
\begin{center}
\renewcommand{\baselinestretch}{1.18}
%{\small
\begin{tabular}{|ll|c|c|c|c|c|c|c|c|c|}
\hline
\multicolumn{2}{|c|}{liquid}      &$R_{\rm WS}$
                                       &$\sigma_\infty$
                                             &$\tau_{\rm o-Ps}$
                                                  &$\Theta_{\rm FWHM}$
                                                           &$R_\infty$
                                                                &$\delta$
                                                                    &$R$
                                 &${\displaystyle \frac{\tilde\sigma}{\sigma_\infty}}$
                                 &${\displaystyle \frac{2\Delta}{R_{\rm WS}}}$ \\
                  &              &\AA&${\displaystyle \rm \frac{dyn}{cm^2}}$
                                             & ns &mrad    & \AA&\AA&\AA &    &   \\
\hline
n-C$_5$H$_{12}$   & n-pentane    &3.583&15.32&4.25&2.25    & 7.4&2.0& 5.4&0.43&2.0\\
n-C$_6$H$_{14}$   & n-hexane     &3.737&17.74&3.92&2.26    & 7.4&2.1& 5.3&0.38&2.3\\
n-C$_7$H$_{16}$   & n-heptane    &3.881&19.65&3.85&2.33    & 7.1&2.0& 5.1&0.37&2.2\\
n-C$_{10}$H$_{22}$& n-decane     &4.267&23.3 &3.49&2.47    & 6.7&2.0& 4.7&0.38&1.8\\
n-C$_{12}$H$_{26}$& n-dodecane   &4.492&24.84&3.43&2.43    & 6.8&2.0& 4.8&0.33&2.2\\
n-C$_{14}$H$_{30}$&n-tetradecane &4.696&26.0 &3.35&2.52    & 6.6&2.0& 4.6&0.35&1.9\\
i-C$_8$H$_{18}$   & isooctane    &4.038&18.33&4.05&2.42    & 6.9&1.9& 5.0&0.45&1.5\\
C(CH$_3$)$_4$     & neopentane   &3.656&11.52&5.15&2.21    & 7.5&1.9& 5.6&0.53&1.4\\
C$_6$H$_{12}$     & cyclohexane  &3.509&24.65&3.24&2.45    & 6.8&2.1& 4.7&0.38&2.2\\
C$_7$H$_{14}$ & methylcyclohexane&3.692&23.28&3.50&2.49    & 6.7&2.0& 4.7&0.41&1.8\\
C$_6$H$_6$        & benzene      &3.285&28.22&3.15&2.55    & 6.5&2.0& 4.5&0.39&2.1\\
C$_6$H$_5$CH$_3$  & toluene      &3.486&27.92&3.24&2.57    & 6.5&2.0& 4.5&0.40&2.0\\
C$_2$H$_5$C$_6$H$_5$&ethylbenzene&3.654&28.74&3.02&2.44    & 6.8&2.1& 4.7&0.32&2.8\\
(CH$_3$)$_2$C$_6$H$_4$& o-xylene &3.769&29.76&3.08&2.54    & 6.5&2.0& 4.5&0.35&2.3\\
(CH$_3$)$_2$C$_6$H$_4$& m-xylene &3.658&28.47&3.20&2.49    & 6.7&2.0& 4.7&0.34&2.4\\
(CH$_3$)$_2$C$_6$H$_4$& p-xylene &3.663&28.01&3.21&2.49    & 6.7&2.0& 4.7&0.35&2.4\\
(CH$_3$)$_3$C$_6$H$_3$&mesitylene&3.748&27.55&3.21&2.49    & 6.7&2.0& 4.7&0.35&2.3\\
(CH$_3$)$_4$C$_6$H$_2$&1,2,3,4-thetra-
                                 &3.888&29.  &3.02&2.52    & 6.6&2.0&  4.6&0.34&2.2\\
           & methylbenzene       &     &     &    &        &    &   &    &    &   \\
C$_6$F$_6$ & hexafluorobenzene   &3.572&22.63&3.78&2.39    & 7.0&2.0& 5.0&0.37&2.4\\
Si(CH$_3$)$_4$&tetramethylsilane &3.780&13.20&4.75&2.25    & 7.4&1.9& 5.5&0.49&1.5\\
(C$_2$H$_5)_2$O & diethylether   &3.453&16.65&3.82&2.29    & 7.6&2.1& 5.2&0.44&1.9\\
 1,4-C$_4$H$_8$O$_2$ & dioxane   &3.234&32.61&3.02&2.86    & 5.8&1.8& 4.0&0.52&1.1\\
 CH$_3$OH            & methanol  &2.528&22.12&3.58&2.29    & 7.2&2.1& 5.2&0.37&3.5\\
 C$_2$H$_5$OH        & ethanol   &2.855&21.97&3.50&2.35    & 7.1&2.1& 5.0&0.39&2.7\\
 C$_3$H$_7$OH        & propanol  &3.101&23.32&3.38&2.40    & 6.9&2.0& 4.9&0.39&2.5\\
 C$_4$H$_9$OH        & butanol   &3.311&24.93&3.36&2.46    & 6.7&2.0& 4.7&0.39&2.3\\
 C$_8$H$_{17}$OH     & octanol   &3.938&27.10&3.13&2.57    & 6.5&2.0& 4.5&0.39&1.8\\
 H$_2$O              & water     &1.928&72.14&1.85&3.05    & 5.4&2.0& 3.4&0.39&2.8\\
 D$_2$O              &heavy water&1.930&70.89&1.95&2.87    & 5.8&2.1& 3.7&0.31&4.1\\
(CH$_3)_2$CO         & acetone   &3.078&24.02&3.29&2.45    & 6.8&2.0& 4.8&0.41&2.2\\
 CH$_3$CN          & acetonitrile&2.756&28.66&3.30&2.44    & 6.8&2.0& 4.8&0.35&3.2\\
 CS$_2$        & carbon disulfide&2.880&31.58&2.20&2.35    & 7.1&2.5& 4.6&0.29&3.9\\
 He, 4.2 K           & helium    &2.350&0.096&99.1&0.86(6) &19.3&1.9&17.4&0.97$^{+0.6}_{-0.5}$  &0.2$^{+9}_{-2} $\\
 H$_2$, 20.3 K       & hydrogen  &2.243& 1.92&28.6&1.3(2)  &12.8&2.1&10.7&0.44$^{+0.34}_{-0.22}$&6.1$^{+14}_{-5}$\\
 N$_2$, 77.3 K       & nitrogen  &2.395& 8.85&11.0&1.8(2)  & 9.2&1.8& 7.4&0.35$^{+0.17}_{-0.13}$&5.7$^{+6}_{-4} $\\
 Ar,    86.4 K       & argon     &2.245&12.42&6.50&2.20(15)& 7.6&1.8& 5.8&0.56$^{+0.15}_{-0.13}$&2.0$^{+2}_{-1} $\\
\hline
\end{tabular}
%}
\end{center}
\medskip
$R_{\rm WS}$ is the radius of the Wigner-Seitz cell (calculated from the
density and molecular mass).\\

$\sigma_\infty$ is the surface tension coefficient of a liquid with plane
interphase boundary at room temperature (except the cases of liquified
gases).\\

$\tau_{\rm o-Ps}$ is the o-Ps lifetime \cite{Nak88,Ste90}. \\

$\Theta_{\rm FWHM}$ is the full width at half maximum of the narrow
component of the ACAR spectra \cite{Nak88,Ste90}. \\

$R_\infty = R+\delta$ is the radius of the Ps bubble, calculated from
Eq.(3.21) using $\Theta_{\rm FWHM}$ values.\\

$\delta$ is the penetration depth of the outer electrons into the Ps bubble.
It is obtained from Eq.(3.9) using $R_\infty$ and $\tau_{\rm o-Ps}$
values.  $\tau^0_{\rm o-Ps}$ was adopted to be 0.5 ns in all cases except He
(1.9 ns), H$_2$ (0.92 ns) and N$_2$ (0.56 ns).\\

Relative uncertainties of $\tau_{\rm o-Ps}$ and $\Theta_{\rm FWHM}$ are no
more than 5 and 10\% respectively (for liquified gases indicated in
parenthesis). Uncertainty of $\tilde\sigma/\sigma_\infty$ is approximately
four times larger than uncertainties of $\Theta_{\rm FWHM}$
(about 40\%).\\

$2\Delta/R_{\rm WS}$ are calculated from Eq.(1.1) using respective values
of $\tilde\sigma/\sigma_\infty$ and assuming $r\equiv R$ (in accord with
Eq.(4.9)).

\end{table*}

%%%%%%%%%%%%%%%%%%%%%%%%%%%%%%%%%%%%%%%%%%%%%%%%%%%%%%%%%%%%%%%%%%%%%%%
%\clearpage
\begin{table*}[t]
\caption{Parameters of the Ps trap and microscopic surface tension
         obtained within the finite potential well model.}
\medskip
\begin{center}
\renewcommand{\baselinestretch}{1.15}
%{\normalsize
\begin{tabular}{|rl|c|c|c|c|c|c|c|c|c|c|}
\hline
  & liquid          &$R_U$&$\delta$
                             &$R$ &$U$ &$E$ &$k_UR_U$
                                                 &${\displaystyle \frac{\tilde\sigma}{\sigma_\infty}}$
                                                 &$\tilde\sigma(R_\sigma)$
                                                 &${\displaystyle \frac{2\Delta}{R_{\rm WS}}}$
                                                 &$p_{cr}$\\
  &                     & \AA&\AA& \AA& eV & eV &    &    &${\displaystyle \rm \frac{dyn}{cm^2}}$ && MPa\\
\hline
1 & n-pentane           & 6.4&1.2& 5.2&3.48&0.37&2.81&0.49& 7.45 &1.5&3.36\\
2 & n-hexane            & 6.3&1.2& 5.1&3.32&0.38&2.80&0.43& 7.56 &1.8&3.01\\
3 & n-heptane           & 6.1&1.2& 5.0&3.49&0.40&2.80&0.42& 8.31 &1.8&2.76\\
4 & n-decane            & 5.7&1.1& 4.6&3.66&0.45&2.78&0.43& 9.99 &1.4&2.10\\
5 & n-dodecane          & 5.8&1.2& 4.6&3.52&0.44&2.78&0.37& 9.24 &1.8&1.83\\
6 & n-tetradecane       & 5.6&1.1& 4.5&3.72&0.47&2.78&0.40&10.3  &1.4&1.61\\
7 & isooctane           & 5.9&1.1& 4.8&3.91&0.43&2.80&0.51& 9.27 &1.1&2.57\\
8 & neopentane          & 6.6&1.1& 5.5&3.84&0.35&2.83&0.58& 6.71 &1.1&3.37\\
9 & cyclohexane         & 5.7&1.2& 4.5&3.43&0.45&2.77&0.44&10.8  &1.7&4.07\\
10& methylcyclohexane   & 5.7&1.1& 4.6&3.74&0.46&2.78&0.47&11.0  &1.4&3.47\\
11& benzene             & 5.5&1.1& 4.4&3.64&0.48&2.77&0.46&12.9  &1.6&4.90\\
12& toluene             & 5.5&1.1& 4.4&3.77&0.49&2.77&0.46&12.8  &1.5&4.10\\
13& ethylebenzene       & 5.7&1.2& 4.5&3.25&0.45&2.76&0.37&10.6  &2.1&3.60\\
14& o-xylene            & 5.5&1.2& 4.3&3.57&0.48&2.77&0.40&12.0  &1.7&3.73\\
15& m-xylene            & 5.6&1.2& 4.4&3.52&0.46&2.77&0.40&11.3  &1.8&3.54\\
16& p-xylene            & 5.6&1.1& 4.5&3.52&0.46&2.77&0.40&11.2  &1.8&3.51\\
17& mesitylene          & 5.6&1.1& 4.5&3.52&0.46&2.77&0.40&11.1  &1.8&3.13\\
18&1,2,3,4-thetra-      & 5.5&1.2& 4.3&3.46&0.48&2.76&0.40&11.6  &1.7&2.9 \\
  & methylbenzene       &    &   &    &    &    &    &    &      &   &    \\
19& hexafluorobenzene   & 5.9&1.1& 4.8&3.64&0.42&2.79&0.42& 9.47 &1.9&3.27\\
20& tetramethylsilane   & 6.4&1.1& 5.3&3.77&0.37&2.82&0.54& 7.15 &1.2&2.82\\
21& diethylether        & 6.2&1.2& 5.0&3.36&0.39&2.80&0.49& 8.23 &1.5&3.64\\
22& dioxane             & 4.9&1.0& 3.9&4.46&0.61&2.76&0.60&19.7  &0.8&4.07\\
23& methanol            & 6.2&1.2& 5.0&3.21&0.39&2.79&0.42& 9.37 &2.7&8.09\\
24& ethanol             & 6.0&1.2& 4.8&3.33&0.41&2.78&0.45& 9.91 &2.1&6.13\\
25& propanol            & 5.9&1.2& 4.7&3.39&0.43&2.78&0.45&10.4  &1.9&5.17\\
26& butanol             & 5.7&1.1& 4.6&3.54&0.45&2.78&0.45&11.1  &1.7&4.42\\
27& octanol             & 5.4&1.1& 4.3&3.69&0.49&2.77&0.45&12.2  &1.3&2.86\\
28& water               & 4.3&1.1& 3.2&3.64&0.73&2.68&0.49&35.6  &1.7&22.1\\
29& heavy water         & 4.6&1.2& 3.4&3.33&0.64&2.69&0.39&27.9  &2.8&    \\
30& acetone             & 5.8&1.2& 4.6&3.46&0.45&2.77&0.47&11.3  &1.7&5.27\\
31& acetonitrile        & 5.8&1.2& 4.6&3.44&0.44&2.77&0.41&11.7  &2.4&4.85\\
32& carbon disulfide    & 5.7&1.4& 4.3&2.43&0.43&2.71&0.35&11.2  &2.8&7.90\\
  & He, 4.2 K           &18.6&1.1&17.5&3.74&0.05&3.02&0.98$^{+0.6}_{-0.5}$  &0.094&0.1$^{+8}_{-2}$  &0.23\\
  & H$_2$, 20.3 K       &11.9&1.2&10.7&3.16&0.12&2.95&0.46$^{+0.35}_{-0.23}$& 0.88&5.7$^{+13}_{-4}$ &1.29\\
  & N$_2$, 77.3 K       & 8.4&1.1& 7.3&4.05&0.23&2.90&0.37$^{+0.18}_{-0.13}$& 3.3 &5.2$^{+6}_{-3}$  &3.39\\
  & Ar,    86.4 K       & 6.7&1.0& 5.7&4.50&0.35&2.86&0.60$^{+0.17}_{-0.14}$& 7.5 &1.7$^{+1.4}_{-1}$&4.90\\
\hline
\end{tabular}
%}
\end{center}

Calculating parameters of the Ps bubble we adopted that $\pi r_0^2
c Z_{\rm eff}n$ is approximately equal to 0.5 ns in all cases except He (1.9
ns), H$_2$ (0.92 ns) and N$_2$ (0.56 ns). \\

$2\Delta/R_{\rm WS}$ are calculated from Eq.(1.1) using respective values
of $\tilde\sigma/\sigma_\infty$ and assuming $r\equiv R$ (in accord with
Eq.(4.9)).

\end{table*}

\end{document}